# Manufacturing of Textured Bulk Fe-SmCo$_5$ Magnets by Severe Plastic Deformation

**Lukas Weissitsch [1],\*, Martin Stückler [1], Stefan Wurster [1], Juraj Todt [2], Peter Knoll [3], Heinz Krenn [3], Reinhard Pippan [1] and Andrea Bachmaier [1]**

[1]Erich Schmid Institute of Materials Science of the Austrian Academy of Sciences, 8700 Leoben, Austria;
[2]Department of Material Physics, Montanuniversität Leoben, 8700 Leoben, Austria;
[3]Institute of Physics, University of Graz, 8010 Graz, Austria;
**\***Correspondence: lukas.weissitsch@oeaw.ac.at

**Abstract:** Exchange-coupling between soft- and hard-magnetic phases plays an important role in the engineering of novel magnetic materials. To achieve exchange coupling, a two-phase microstructure is necessary. This interface effect is further enhanced if both phase dimensions are reduced to the nanometer scale. At the same time, it is challenging to obtain large sample dimensions. In this study, powder blends and ball-milled powder blends of Fe-SmCo$_5$ are consolidated and are deformed by high-pressure torsion (HPT), as this technique allows us to produce bulk magnetic materials of reasonable sizes. Additionally, the effect of severe deformation by ball-milling and severe plastic deformation by HPT on exchange coupling in Fe-SmCo$_5$ composites is investigated. Due to the applied shear deformation, it is possible to obtain a texture in both phases, resulting in an anisotropic magnetic behavior and an improved magnetic performance.

## 1. Introduction

One parameter, which limits the performance of permanent magnets, is a sub-optimal remanent magnetization ($M_r$). This usually leads to a reduced energy product—$BH_{max}$—albeit only if a very high coercivity ($H_c$) is reached. An enhancement can generally be realized by three pathways. First, manufacturing a textured magnetic material, where the magnetically easy axis of many crystallites is aligned in one direction. Such textured magnets show a pronounced rectangular hysteresis loop when the measuring field is applied parallel to the easy axis and are referred to as anisotropic magnets [1]. Second, by combining a hard-magnetic with a soft-magnetic phase forming an exchange-spring magnetic material. If the soft-magnetic phase is finely dispersed within the hard phase, the high magnetization of the soft-magnetic material is stabilized by the large anisotropy field of the hard-magnetic material, as initially described by Kneller and Hawig [2]. Therefore, the magnetic polarization of the soft phase remains oriented with the enclosing hard phase resisting demagnetization, while it increases the magnetic moment of the entire material. Third, a refinement of the microstructure could lead to an increase in $M_r$. In addition, a unimodal grain size distribution increases the coercivity [3,4].

Combining all three pathways and simultaneously manufacturing bulk-like magnets, which are indispensable for large-scale applications, is challenging [5]. Nanocrystalline microstructures can be achieved, e.g., by ball-milling and the ball-milled powders can be aligned in external magnetic fields to obtain a texture [6,7]. However, subsequent powder consolidation usually involves thermal treatments, which leads to grain growth and a coarsening of the microstructure [8].

In a previous study, we already demonstrated the feasibility of severe plastic deformation (SPD) using HPT-deformation of Fe and $SmCo_5$ powders to produce Fe-$SmCo_5$ exchange-coupled spring magnets with a broad range of chemical compositions [9]. Therein, special emphasis was devoted to the evolution of the microstructure during deformation and its correlation to magnetic properties. It was possible to reduce grain sizes and phase dimensions by HPT-deformation, while simultaneously being able to obtain bulk-sized magnets. A nanocrystalline, lamellar, dual phase structure with phase thicknesses below 1µm was presented for deformation at room temperature. Additionally, an increasing saturation magnetization ($M_s$) and $H_c$ with an increasing deformation strain was observed. However, the maximum $H_c$ already peaked at a relatively low amount of applied strain. A beneficial effect of plastic deformation on the magnetic performance is also reported for different material systems, which could peak at a certain amount of applied strain [10,11].

Herein, we demonstrate the combination of all three approaches—texture, exchange-coupling, phase and grain refinement—to improve the performance of permanent magnets by SPD using HPT. The study is performed in direct comparison with ball-milled and consolidated Fe and $SmCo_5$ powders. Thus, the effect of HPT-deformation on the magnetic properties can be isolated. The ball-milled powders are then subsequently HPT-deformed, which allows a study of the effect of preceding ball-milling on the HPT-deformation and the resulting magnetic properties. Besides, the potential of HPT-deformation to induce texture is proven.

## 2. Materials and Methods

Powder blends with three chemical compositions (66, 43 and 26 wt.% Fe: MaTeck 99.9% – 100 + 200 mesh and remaining $SmCo_5$: Alfa Aesar, intermetallic fine powder; in the following denoted with the suffix 1–3, respectively) are used for the experiments. Parts of the powder blends are milled in an air-cooled planetary ball mill (Retsch PM400) with an 1:15 powder-to-ball ratio of 400 rpm between 1–3 h. To prevent heat development inside the jar, an alternating 5 minutes' interval of mill-time and cooling-time was programmed. Powder handling, ball-milling and the consolidation process are carried out in an Ar-atmosphere. All powder blends (un-milled and ball-milled) are hydrostatically consolidated using an HPT device. Parts of the consolidated pellets are subsequently HPT-deformed by applying 7.5 GPa at room temperature. Thus, three different types of samples are obtained: ball-milled powder samples (denoted as 'BM' in the following), ball-milled and HPT deformed powder samples ('BM + HPT') and un-milled but HPT-deformed powder samples ('HPT'). Based on our previous work [9], we limited HPT-deformation for this study to a maximum number of 3 rotations, which corresponds to a shear strain γ = 56 at a radius of 3 mm. HPT disc sizes are 8 mm in diameter and about 0.6 mm in thickness, whereas samples for magnetic measurements were cut out with a wire saw. In Figure 1, the positions of measurements and measurement directions in relation to HPT-deformation directions are given.

Demagnetizing curves (1st half of complete hysteresis curve) between ±70 kOe at 300 K, starting from $H_{max}$ = +70 kOe, are recorded with a SQUID-Magnetometer (Quantum Design MPMSXL-7, Quantum Design, Inc., San Diego, CA, USA) using the manufacturer's software MPMSMultiVu Application (version 1.54). The magnetic field was applied in different HPT-disc directions as depicted in Figure 1.

To investigate the microstructure after HPT-deformation, scanning electron microscopy (SEM; LEO 1525, Carl Zeiss Microscopy GmbH, Oberkochen, Germany) in backscattered electron detection mode (BSE) is performed on the HPT-deformed samples. For texture analysis, electron backscatter diffraction (EBSD) measurements are performed with a Bruker e-Flash[FS] detector on an HPT-deformed Fe sample.

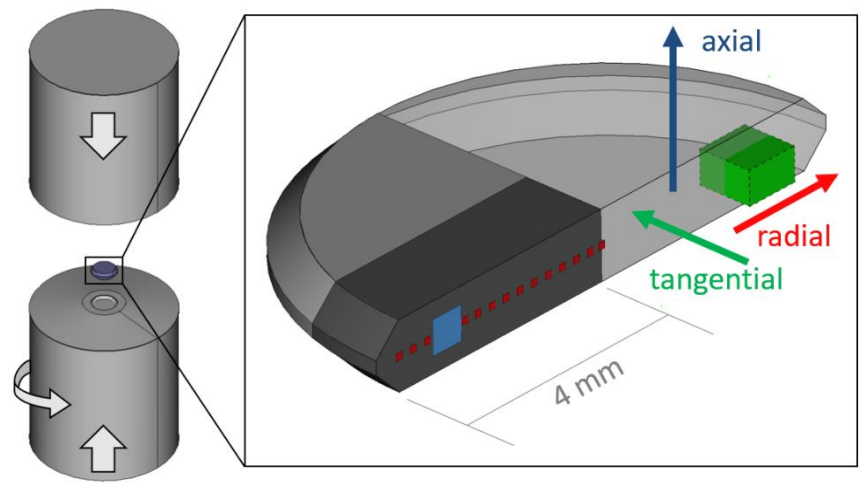

**Figure 1**. Schematic representation of the HPT-setup consisting of two anvils and the sample whereat a halved HPT-disc is shown in more detail. The position and size of a SQUID specimen is highlighted in green. Respective HPT-deformation directions are shown as coloured arrows. SEM measurements are conducted in a tangential direction at a radius ≈ 3 mm as highlighted in blue. Radially resolved HEXRD texture analysis is performed on a specimen (dark grey) in a tangential direction (red areas) in transmission mode.

Commercially available and ball-milled powder blends were characterized using conventional X-ray diffraction (XRD, D2 Phaser Bruker Corporation, Billerica, MA, USA) with Co-$K_\alpha$ radiation after different milling times. The theoretical reference peaks were found in the Crystallography Open Database (SmCo$_5$: COD 1524132, Fe: COD 9006587) [12].

Synchrotron X-ray diffraction (HEXRD) measurements in transmission mode (Petra III: P07B, DESY, Hamburg, Germany) were carried out to analyze the texture selected nanocomposite samples (with highest SmCo$_5$ content) as a function of the radius. The beam energy used was 87.1 keV and the shutter size 0.1 × 0.1 mm$^2$, while using a Perkin Elmer XRD 1621 detector. The obtained data were calibrated with a diffraction pattern of LaB$_6$ powder (LaB$_6$: COD 1000057) and further processed with DigitalMicrograph using the PASAD package (PASAD-tools.v0.1) [13] and an in-house-developed

Matlab[TM] code. The codes' output gives an integrated intensity as a function of the angle corresponding to the Debye diffraction cone for every chosen diffraction peak. For pole figure analysis and orientation density function (ODF) estimations, the Matlab[TM] toolbox MTEX (version 5.7.0) was used [14]. Therein, the selected and processed Debye ring was used to extrapolate an ODF estimation, allowing us to recalculate a pole figure. Due to the high crystal symmetry of the investigated phases, a full ODF by spherical tomography is not necessary to claim on a certain present texture [15]. However, large areas on the pole figure representations are extrapolated and likely contain errors. This is why a transparent mask is used to overlay the pole figure representations in this work, indicating a focus on the information obtained from uncovered areas.

## 3. Results and Discussion

### 3.1. Sample Processing and Magnetic Properties

XRD measurements of commercially available and ball-milled powder blends, containing 34, 57 and 74 wt.% $SmCo_5$ and remaining Fe, are presented in Figure 2a–c, respectively. In the initial state (bottom pattern), Fe and $SmCo_5$ peaks are visible for all compositions. To monitor the ball-milling process, parts of the powder blends are removed in steps of 30 min and are measured by XRD. With the increasing milling time, a significant peak broadening is observed for both phases. The broadening of the $SmCo_5$ peaks takes place already at an earlier stage of ball-milling and therefore the peaks tend to vanish for longer milling times and for powder blends with lower $SmCo_5$ contents. Hence, a partly amorphous phase cannot be excluded. The powder blend with the highest $SmCo_5$ content (Figure 2c) tends to agglomerate and due to the similarity of the 30 min and 60 min XRD pattern, no further energy input into the powder blend by prolonged ball milling is expected.

In the literature, an intermixing of elements, especially the formation of α-Fe(Co) and interstitial SmCo(Fe) phases, are reported [16–19]. Although the formation of other phases seems to originate in the severe deformation process, an additionally temperature process is present in all reports. In this study, a temperature treatment of the powders as well as during the HPT-processing is completely evaded. However, we do not exclude the formation or diffusion of Fe or Co on an atomic scale, but in all recorded XRD patterns, no additional peaks are found. Thus, neither decomposition of $SmCo_5$ nor a significant formation of an FeCo phase during milling is observed.

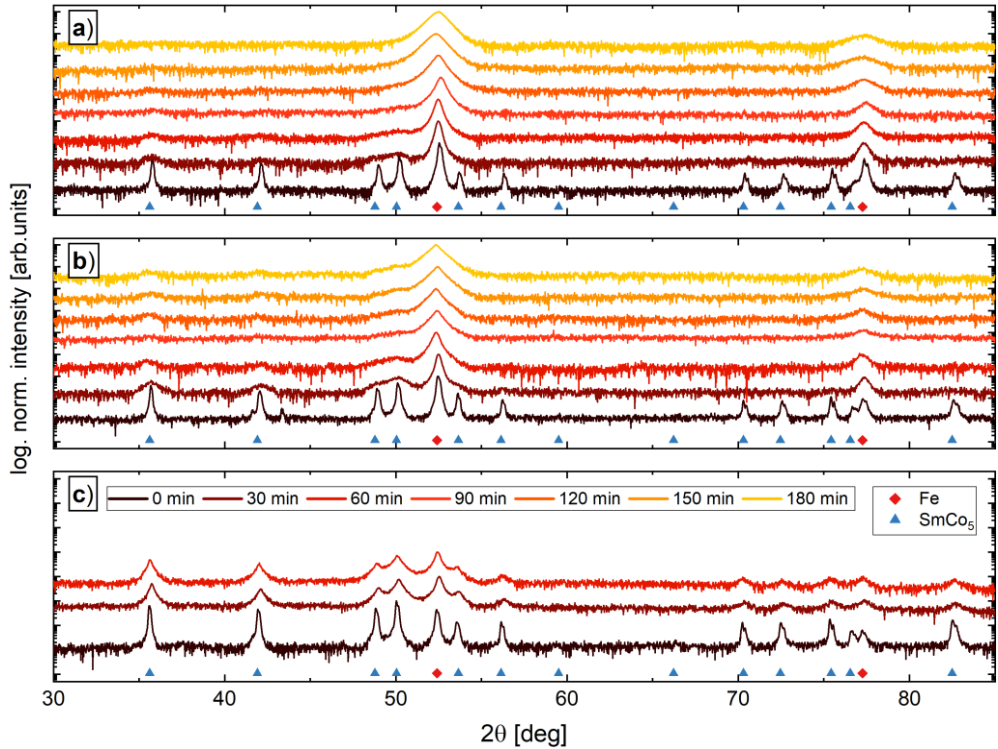

**Figure 2.** XRD measurements as a function of milling time for powders consisting of 66 wt.% Fe—34 wt.% SmCo$_5$ (a), 43 wt.% Fe—57 wt.% SmCo$_5$ (b) and 26 wt.% Fe—74 wt.% SmCo$_5$ (c). The theoretical reference peaks are found in the Crystallography Open Database (SmCo$_5$: COD 1524132, Fe: COD 9006587).

In Figure 3, the demagnetizing curves of the samples with the highest SmCo$_5$ content are exemplarily shown for the un-milled powder blend and all processing routes. The processing history has a strong influence on the demagnetization curve. Ball-milling barely increases H$_c$, but leads to an increased M$_s$. HPT-deformation of the BM powder decreases the susceptibility, while M$_r$ remains constant but M$_s$ and H$_c$ increases. Besides a slightly smaller M$_s$, this behavior is even more strongly pronounced for the 'HPT' sample.

The magnetic properties of all processed samples are summarized in Table 1. H$_c$ is determined by a linear fit of data points close to zero crossing. An increase of M$_r$ and H$_c$ with increasing SmCo$_5$-content is found for all samples. For all compositions, both magnetic parameters are highest for HPT-deformed samples without previous ball-milling.

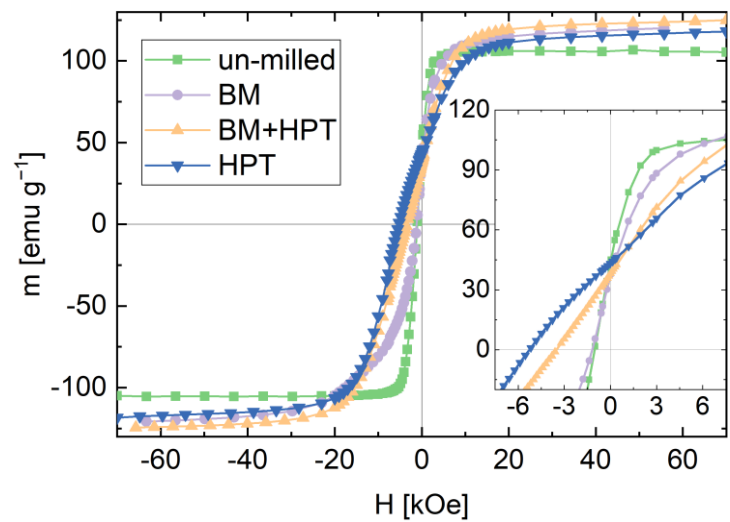

**Figure 3**. Demagnetizing curves starting from +70 kOe for the un-milled powder blend, ball-milled powder sample (BM), ball-milled and HPT-deformed powder sample (BM + HPT) and un-milled and HPT-deformed powder sample (HPT), exemplified for samples with the highest $SmCo_5$-content. The inset shows the demagnetizing branches in more detail.

The approach to the presented $M_s$ value is obtained by a linear fit for 1/H at fields above 35 kOe, determining the intersection with the ordinate. Samples containing a higher amount of Fe show a higher $M_s$. Expected $M_s$ values for single phase $SmCo_5$ are 112 emug$^{-1}$ [20] and 218 emug$^{-1}$ for bulk Fe [21]. Applying a superposition based on the samples' chemical composition depicted in Table 1, a calculated $M_s$ reveals 180, 158 and 138 emug$^{-1}$ for composition 1, 2 and 3, respectively. $M_s$ values for HPT-deformed samples lay constantly beneath these theoretical values.

**Table 1**. Magnetic properties measured by SQUID magnetometry. Ball-milled and powder samples are denoted as 'BM', ball-milled and HPT deformed powder samples as 'BM + HPT' and un-milled and HPT deformed powder samples as 'HPT'. The prefix 'rad' and 'tang' denotes measurements with the magnetic field applied parallel to the radial or tangential HPT-disc direction, respectively. For all other listed HPT-deformed samples without any prefix, the field is applied parallel to the axial HPT-disc direction. The suffix 1–3 corresponds to the chemical composition. $H_c$ depicts the coercivity measured in Oe, $M_s$ is the saturation magnetization determined as described in the text and $M_r$ denotes the remanence, both given in the specific mass moment emu g$^{-1}$.

| Sample | Fe/(wt.%) | $SmCo_5$/(wt.%) | $H_c$/(Oe) | $M_s$/(emu g$^{-1}$) | $M_r$/(emu g$^{-1}$) |
|---|---|---|---|---|---|
| un-milled3 | 26 | 74 | −1033 | 106 | 43.2 |
| BM3 | 26 | 74 | −1184 | 125 | 36.4 |
| BM + HPT1 | 66 | 34 | −274 | 196 | 5.4 |
| BM + HPT2 | 43 | 57 | −686 | 172 | 10.2 |
| BM + HPT3 | 26 | 74 | −3526 | 128 | 38.0 |
| rad BM + HPT3 | 26 | 74 | −1472 | 128 | 28.9 |
| tang BM + HPT3 | 26 | 74 | −1434 | 129 | 36.5 |
| HPT1 | 66 | 34 | −887 | 172 | 14.1 |
| HPT2 | 43 | 57 | −2366 | 150 | 30.8 |
| HPT3 | 26 | 74 | −5264 | 122 | 43.2 |
| rad HPT3 | 26 | 74 | −1192 | 120 | 26.3 |
| tang HPT3 | 26 | 74 | −1903 | 120 | 35.3 |

Due to the domain-wall pinning mechanisms and/or a present texture, one could argue that even 70 kOe are not sufficient enough to completely saturate the material [22,23]. For the ball-milled and HPT-deformed samples, on the other hand, the $M_s$ values are partly above the theoretical value. This is most likely due to processing steps itself, i.e., ball-milling, where abrasions of the hardened steel jar and balls, but also powder-handling could enrich the Fe content of the powders. However, the powder blends containing the highest amount of $SmCo_5$ could only be milled for 1 h and the increase in $M_s$ is more strongly pronounced for powders with a lower $SmCo_5$ content which have been milled for a longer time (3 h). Thus, the assumption of Fe enrichment within the ball-milling and powder-handling process is further supported.

To enable exchange coupling, the magnetic materials should exhibit a two-phase microstructure with alternating hard- and soft-magnetic phases. A simple consideration, valid for a perfect model system, implemented by a length within a complete exchange coupling, is expected. This critical length for the soft magnetic phase is approximately defined by twice the Bloch-wall width $\delta_w = \pi \sqrt{\frac{A^{hard}}{K^{hard}}}$ of the hard phase, as its magnetic stray field only reaches a certain distance and hinders the soft phase to change its orientation upon an opposed magnetic field. Here, $A^{hard}$ and $K^{hard}$ are the exchange and anisotropy constant of the hard-magnetic phase [24,25]. With $A^{hard}$ = 1.2 × 10$^{-11}$ Jm$^{-1}$ and $K^{hard}$ = 17.2 × 10$^6$ Jm$^{-3}$ [26,27] for SmCo$_5$, an optimum soft phase thickness of ≈ 5.3 nm is obtained. Another approximation for the soft phase thickness is based on the idea of how easy the spins of that material align with the hard phase. The exchange length $l_{ex}^{soft} = \sqrt{\frac{A^{soft}}{K^{soft}}}$ depends on $A^{soft}$ and $K^{soft}$, expressing the same constants but for the soft-magnetic phase. For α-Fe: $A^{soft}$ = 2.5 × 10$^{-11}$ Jm$^{-1}$ and $K^{soft}$ = 4.6 × 10$^4$ Jm$^{-3}$, $l_{ex}$ ≈ 23 nm is obtained [2,24]. Although this soft phase thickness differs by a factor of ≈ 4, both values give an idea for the optimum dimensions of the Fe phase.

To investigate the microstructure after HPT-deformation, scanning electron microscopy in backscattered electron detection mode (BSE) is performed on the HPT-deformed samples (Figure 4). Using a BSE contrast, Fe or Fe-rich areas exhibit a dark contrast and SmCo$_5$ or SmCo$_5$-rich areas a bright contrast, respectively. Although the same chemical composition is present, the microstructure of the samples with the lowest SmCo$_5$ content after HPT-deformation differs significantly. Ball-milling has a clear influence on the dispersion of the phases, as contrast differences in Figure 4a are hardly visible, while a more heterogeneous microstructure is depicted in Figure 4d for the HPT sample. For both samples, however, a lamellar microstructure, well suited for exchange coupling, is missing. With the increasing SmCo$_5$ content, a similar morphology is obtained for the BM + HPT2 sample (Figure 4b). Although, the larger SmCo$_5$ phases are visible as well. In the case of the 'HPT'-sample, a lamellar morphology is obtained (Figure 4e). A similar lamellar microstructure is found for the samples with the highest SmCo$_5$ content (Figure 4c,f). For these samples, different thicknesses of the phases are present, and all are well below 1μm. For exchange coupling, theoretical considerations prefer the soft-magnetic phase embedded in a hard-magnetic matrix [24,28]. Based on XRD and SEM investigations, the HPT-samples (Figure 4e,f) containing 57 and 74 wt.% SmCo$_5$ and the BM + HPT-samples with 74 wt% SmCo$_5$ (Figure 4c) are most promising for a high exchange coupling effect, which is confirmed by the magnetic measurements (cf. Table 1). However, a completely exchanged coupled material is not expected due to the large number of different sized lamellae.

In general, no beneficial influence of initial ball-milling on the microstructural evolution (e.g., homogeneity of deformation to achieve lamellar microstructure for all compositions, prevention of shear band formation, further phase refinement) during HPT-deformation is observed, with consequences for resulting magnetic properties. In fact, a partly amorphous SmCo$_5$ phase resulting from initial ball-milling, which would reduce $H_c$, cannot be excluded.

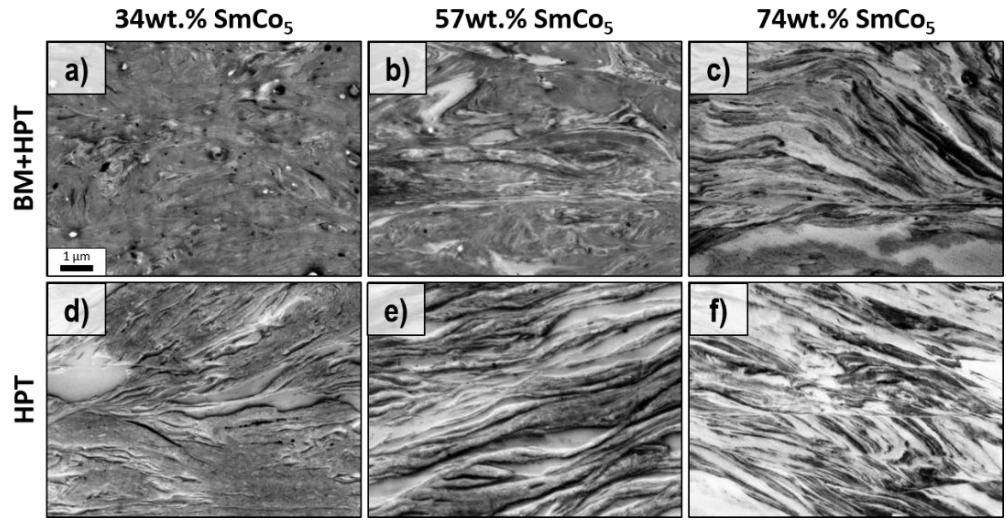

**Figure 4.** BSE images of ball-milled and HPT-deformed powder samples (**a**–**c**) and un-milled and HPT deformed powder samples (**d**–**f**) in tangential direction with 34 (**a**,**d**), 57 (**b**,**e**) and 74 wt.% SmCo$_5$ (**c**,**f**). The scale bar in (**a**) also applies to all other images.

HPT-deformed samples with the highest amount of SmCo$_5$ show the best hard magnetic properties. Thus, magnetic properties of these HPT-deformed samples are measured in other HPT-disc directions as well. In Figure 5, the demagnetization curves in axial (blue), radial (red) and tangential (green) HPT-disc directions are shown for the BM + HPT3 (Figure 5a) and the HPT3 sample (Figure 5b). By applying the magnetic measurement field parallel to the radial and tangential directions drastically reduces H$_c$ for both samples. M$_r$, however, is nearly unchanged. Additionally, the shape of the demagnetizing curves for the radial and tangential HPT-disc directions is different compared to the axial direction and the curves also exhibit an increased susceptibility. Although the characteristics of the radial and tangential curves are relatively similar, the tangential curve saturates faster and resists against demagnetization a little bit longer. Furthermore, the BM + HPT3 sample shows a linear m(H) behavior in the second quadrant, while the HPT sample reveals a convex bent shape. The magnetic properties obtained from these measurements are again summarized in Table 1.

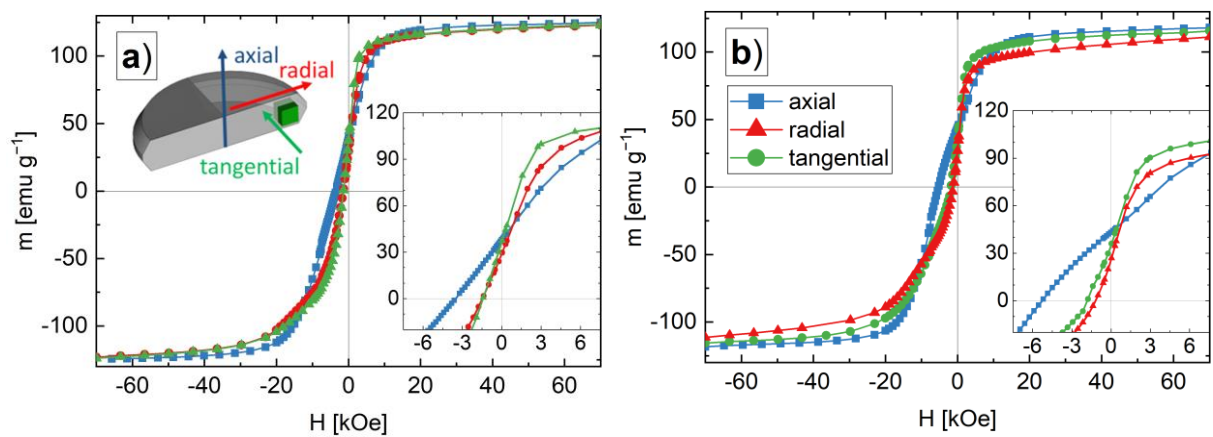

**Figure 5.** Demagnetizing curves starting from +70 kOe for the ball-milled and HPT-deformed powder (BM + HPT3) samples (**a**) and un-milled and HPT-deformed powder (HPT3) samples (**b**). The schematic illustration in (**a**) visualizes the measurement directions in respect of the different HPT-disc directions (axial: blue; radial: red; tangential: green). The insets show the demagnetizing branches in more detail.

For textured hard magnets, the coercivity strongly changes, if the measurement field is applied parallel or perpendicular to the easy axis of the crystals due to its large magnetocrystalline anisotropy [1,29–

31]. The higher susceptibility is caused by the uncoupled volume of the Fe phase, as the magnetic moment originating from $SmCo_5$ in the magnetically hard direction (basal plane) would saturate much slower. The magnetocrystalline anisotropy of Fe is much lower compared with $SmCo_5$, but again, the difference between tangential and radial demagnetization curve can be explained with a textured microstructure. The crystallographic (100) direction is the Fe easy axis, which should be aligned in the tangential HPT-disc direction, whereas the radial direction should correspond to the crystallographic (110) semi-hard direction [29].

*3.2. X-ray Texture Analysis*

Synchrotron X-ray diffraction (HEXRD) measurements are carried out to analyze the texture of the as-deformed BM + HPT3 and HPT3 samples (with highest $SmCo_5$ content) as a function of the radius. The recorded diffraction rings clearly show a variation of intensities, which indicates the existence of a preferred microstructural crystallite orientation [15]. For an appropriate comparison and based on the diffraction rings, pole figure representations are created. To verify our results obtained by synchrotron measurements, EBSD measurements are conducted on an HPT-deformed single-phase Fe sample. In Figure 6a, the EBSD-recorded {110} and {100} pole figures for pure Fe are shown. In Figure 6b–d, the HEXRD Fe {110} and {100} and $SmCo_5$ $\{\bar{2}11\bar{1}\}$ and $\{\bar{2}110\}$ pole figures are presented. The Fe pole figures show the typical simple shear texture for both the EBSD and the synchrotron data-sets, which allows us to extend the pole figure analysis to the $SmCo_5$ phase.

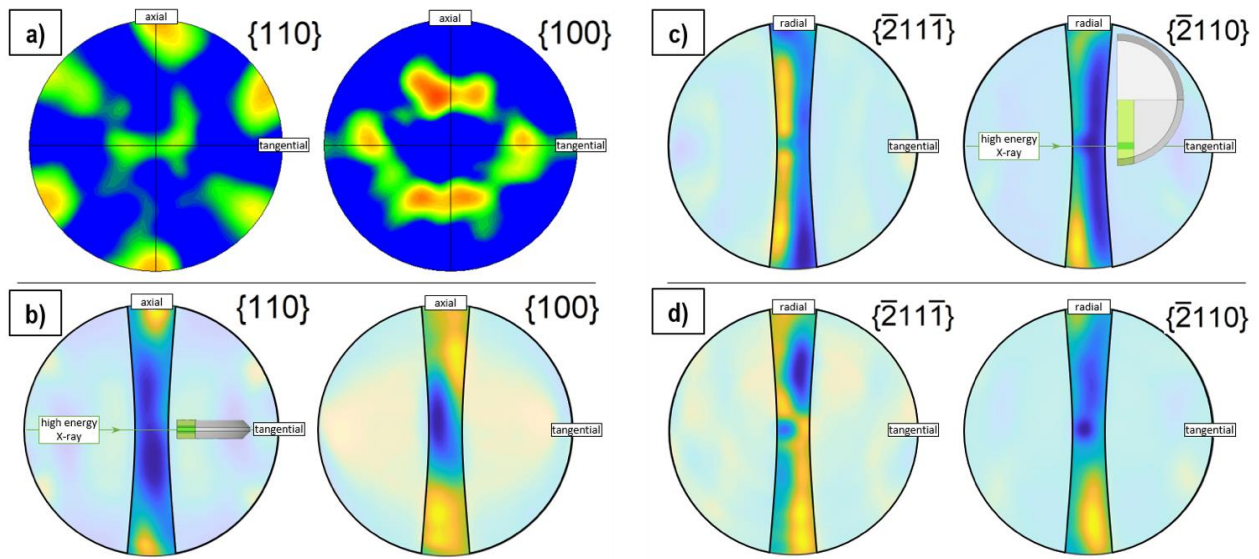

**Figure 6.** EBSD pole figure of Fe based on EBSD measurements (**a**). Pole figure representations of Fe reconstructed based on a HEXRD measurement of the HPT-deformed composite HPT3 (**b**) and a representation of the same sample and diffraction pattern for two $SmCo_5$ planes (**c**). Pole figure representation for the BM + HPT3 sample and the same $SmCo_5$ planes (**d**). An inset in the image (**b**) for the Fe phase and in (**c**) for the $SmCo_5$ phase visualizes the relative position of the HPT disc with respect to the incident X-ray beam and shown pole figures. The position of the HEXRD measurement within the sample is highlighted. The pole figure representations are partly covered with a transparent mask to lay the focus on the information obtained from less extrapolated areas.

The $\{\bar{2}110\}$ planes are oriented parallel to the c-axis with their surface normal in the basal plane of the $SmCo_5$ crystal. The corresponding pole figure shows a minimum in axial HPT-disc direction but a pronounced intensity is visible in radial direction, meaning that the crystal-plane normal preferably pointing in this direction. The $\{\bar{2}11\bar{1}\}$ and $\{\bar{2}110\}$ planes comprise an angle of 39°. For the $\{\bar{2}11\bar{1}\}$ pole figures, this results in an intensity maxima pointing to angles closer to the axial HPT-disc direction. Thus, the hexagonal crystals of the $SmCo_5$ phase tend to align their c-axis parallel to the axial HPT-disc direction with the basal plane parallel to the shear plane, spanned by radial and tangential directions. This behavior is similar for the BM + HPT3 and HPT3 samples, but is strongly pronounced for the latter one and generally for larger applied shear strain (larger radius of HPT-disc). For the hard magnetic $SmCo_5$ phase, a clear correlation of texture with resulting magnetic properties is found. Regarding the soft magnetic Fe phase, the magnetically easy axis directions {100} partly orient in a tangential HPT-disc direction, leading to a stiffer hysteresis loop (cf. Figure 5).

## 4. Conclusions

In summary, HPT deformation offers an easy way to produce exchange-coupled bulk $Fe-SmCo_5$ magnets from commercially available powder blends without any thermal treatments or complicated processing steps. Although ball-milling enhances $H_c$, no beneficial effect for subsequent HPT-deformation is observed compared to un-milled powders. After deformation, the formation of a heterogeneous microstructure with a lamellar morphology and enhanced coercivity is observed. The magnetic anisotropic characteristic of the $Fe-SmCo_5$ magnets is correlated to a crystallographic texture in both the Fe and the $SmCo_5$ phase. The $SmCo_5$ crystals tend to orient their c-axes parallel to the axial HPT-disc direction with their basal plane in the shear plane. The Fe phase aligns its magnetic easy axis in the tangential HPT-disc direction.

**Author Contributions:** Conceptualization, L.W. and A.B.; Methodology, L.W., S.W. and A.B.; Validation, L.W., M.S., S.W., P.K., J.T., H.K., R.P. and A.B.; Formal Analysis, L.W.; Investigation, L.W.; Data Curation, L.W., J.T.; Writing—Original Draft Preparation, L.W.; Writing—Review and Editing, R.P., H.K., P.K., J.T., M.S., S.W., and A.B.; Visualization, L.W.; Supervision, H.K., P.K., R.P. and A.B.; Project Administration, A.B.; Funding Acquisition, A.B. All authors have read and agreed to the published version of the manuscript.

**Funding:** This project has received funding from the European Research Council (ERC) under the EuropeanUnion's Horizon 2020 research and innovation programme (Grant No. 757333).

**Institutional Review Board Statement:** Not applicable.

**Informed Consent Statement:** Not applicable.

**Data Availability Statement:** Not applicable.

**Acknowledgments:** We acknowledge DESY (Hamburg, Germany), a member of the Helmholtz Association HZ Hereon, for the provision of experimental facilities. Parts of this research were carried out at P07B (PETRA III) and we would like to thank Emad Maawad for assistance.

**Conflicts of Interest:** The authors declare no conflict of interest.

## References

1. Hu, D.; Yue, M.; Zuo, J.; Pan, R.; Zhang, D.; Liu, W.; Zhang, J.; Guo, Z.; Li, W. Structure and magnetic properties of bulk anisotropic $SmCo_5/\alpha$-Fe nanocomposite permanent magnets prepared via a bottom up approach. *J. Alloy. Compd.* **2012**, *538*, 173–176. https://doi.org/10.1016/j.jallcom.2012.05.079.

2. Kneller, E.F.; Hawig, R.The exchange-spring magnet: A new material principle for permanent magnets. *IEEE Trans. Magn.* **1991**, *27*, 3588–3560. https://doi.org/10.1109/20.102931.

3. Fischer, R.; Schrefl, T.; Kronmuller, H.; Fidler, J. Grain-size dependence of remanence and coercive field of¨ isotropic nanocrystalline composite permanent magnets. *J. Magn. Magn. Mater.* **1996**, *153*, 35–49. https://doi.org/10.1016/0304-8853(95)00494-7.

4. Shu-Li, H.; Hong-Wei, Z.; Chuan-Bing, R.; Ren-Jie, C.; Bao-Gen, S. Effects of grain size distribution on remanence and coercivity of $Pr_2Fe_{14}B$ nanocrystalline magnet. *Chin. Phys.* **2005**, *14*, 1055. https://doi.org/10.1088/10091963/14/5/036.

5. Yue, M.; Zhang, X.; Liu, J.P. Fabrication of bulk nanostructured permanent magnets with high energy density: Challenges and approaches. *Nanoscale* **2017**, *9*, 3674–3697. https://doi.org/10.1039/c6nr09464c.

6. Rong, C.-B.; Nguyen, V.V.; Liu, J.P. Anisotropic nanostructured magnets by magnetic-field-assisted processing. *J. Appl. Phys.* **2010**, *107*, 09A717. https://doi.org/10.1063/1.3337656.

7. Cui, B.; Gabay, A.; Li, W.; Marinescu, M.; Liu, J.; Hadjipanayis, G. Anisotropic $SmCo_5$ nanoflakes by surfactant assisted high energy ball milling. *J. Appl. Phys.* **2010**, *107*, 09A721. https://doi.org/10.1063/1.3339775.

8. Rong, C.; Zhang, Y.; Poudyal, N.; Xiong, X.; Kramer, M.J.; Liu, J.P. Fabrication of bulk nanocomposite magnets via severe plastic deformation and warm compaction. *Appl. Phys. Lett.* **2010**, *96*, 102513. https://doi.org/10.1063/1.3358390.

9. Weissitsch, L.; Stückler, M.; Wurster, S.; Knoll, P.; Krenn, H.; Pippan, R.; Bachmaier, A. Strain Induced Anisotropic Magnetic Behaviour and Exchange Coupling Effect in $Fe-SmCo_5$ Permanent Magnets Generated by High Pressure Torsion. *Crystals* **2020**, *10*, 1026. https://doi.org/10.3390/cryst10111026.

10. Li, H.; Li, W.; Guo, D.; Zhang, X. Tuning the microstructure and magnetic properties of bulk nanocomposite magnets with large strain deformation. *J. Magn. Magn. Mater.* **2017**, *425*, 84–89. https://doi.org/10.1016/j.jmmm.2016.10.126.

11. Shchetinin, I.V.; Bordyuzhin, I.G.; Sundeev, R.V.; Menushenkov, V.P.; Kamynin, A.V.; Verbetsky, V.N.; Savchenko, A.G. Structure and magnetic properties of $Sm_2Fe_{17}N_x$ alloys after severe plastic deformation by high pressure torsion. *Mater. Lett.* **2020**, *274*, 127993. https://doi.org/10.1016/j.matlet.2020.127993.

12. Gražulis, S.; Daškevič, A.; Merkys, A.; Chateigner, D.; Lutterotti, L.; Quirós, M.; Serebryanaya, R.N.; Moeck, P.; Downs, T.R.; le Bail, A. Crystallography open database (COD): An open-access collection of crystal structures and platform for world-wide collaboration. *Nucleic Acids Res.* **2012**, *40*, D420–D427. https://doi.org/10.1093/nar/gkr900.

13. Gammer, C.; Mangler, C.; Rentenberger, C.; Karnthaler, H. Quantitative local profile analysis of nanomaterials by electron diffraction. *Scr. Mater.* **2010**, *63*, 312–315. https://doi.org/10.1016/j.scriptamat.2010.04.019.

14. Bachmann, F.; Hielscher, R.; Schaeben, H. Texture analysis with MTEX–free and open source software toolbox. *Solid State Phenom.* **2010**, *160*, 63–68. https://doi.org/10.4028/www.scientific.net/SSP.160.63.

15. Wenk, H.-R.; Grigull, S. Synchrotron texture analysis with area detectors. *J. Appl. Crystallogr.* **2003**, *36*, 1040–1049. https://doi.org/10.1107/S0021889803010136.

16. Breton, J.M.L.; Larde, R.; Chiron, H.; Pop, V.; Givord, D.; Isnard, O.; Chicinas, I. A structural investigation´ of $SmCo_5$/Fe nanostructured alloys obtained by high-energy ball milling and

subsequent annealing. *J. Phys. D Appl. Phys.* **2010**, *43*, 085001. https://doi.org/10.1088/0022-3727/43/8/085001.

17. Pop, V.; Dorolti, E.; Vaju, C.; Gautron, E.; Isnard, O.; le Breton, J.-M.; Chicinas, I. Structural and magnetic behaviour of SmCo$_5$/α-Fe nanocomposites obtained by mechanical milling and subsequent annealing. *Rom. Rep. Phys.* **2010**, *55*, 127–136.

18. Shen, Y.; Huang, M.Q.; Turgut, Z.; Lucas, M.S.; Michel, E.; Horwath, J.C. Effect of milling time on magnetic properties and structures of bulk Sm-Co/α-(Fe, Co) nanocomposite magnets. *J. Appl. Phys.* **2012**, *111*, 07B512. https://doi.org/10.1063/1.3673413.

19. Xiong, X.; Rong, C.; Rubanov, S.; Zhang, Y.; Liu, J. Atom probe study on the bulk nanocomposite SmCo/Fe permanent magnet produced by ball-milling and warm compaction. *J. Magn. Magn. Mater.* **2011**, *323*, 2855–2858. https://doi.org/10.1016/j.jmmm.2011.06.035.

20. Foner, S.; McNiff, E., Jr.; Martin, D.; Benz, M. Magnetic properties of cobalt-samarium with a 24-MGOe energy product. *Appl. Phys. Lett.* **1972**, *20*, 447–449. https://doi.org/10.1063/1.1654011.

21. Kin, M.; Kura, H.; Tanaka, M.; Hayashi, Y.; Hasaegawa, J.; Ogawa, T. Improvement of saturation magnetization of Fe nanoparticles by post-annealing in a hydrogen gas atmosphere. *J. Appl. Phys.* **2015**, 117, 17E714. https://doi.org/10.1063/1.4919050.

22. Chau, R.; Maple, M.; Nellis, W. Shock compaction of SmCo$_5$ particles. *J. Appl. Phys.* **1996**, *79*, 9236–9244. https://doi.org/10.1063/1.362598.

23. Zhang, J.; Zhang, S.-Y.; Zhang, H.-W.; Shen, B.-G. Structure, magnetic properties, and coercivity mechanism of nanocomposite SmCo$_5$/α-Fe magnets prepared by mechanical milling. *J. Appl. Phys.* **2001**, *89*, 5601–5605. https://doi.org/10.1063/1.1365430.

24. Skomski, R.; Manchanda, P.; Kumar, P.; Balamurugan, B.; Kashyap, A.; Sellmyer, D.J. Predicting the future of permanent-magnet materials. *IEEE Trans. Magn.* **2013**, *49*, 3215–3220. https://doi.org/10.1109/TMAG.2013.2248139.

25. Blundell, S. *Magnetism in Condensed Matter*; Oxford University Press: New York, NY, USA, 2001.

26. Coey, J.M. *Magnetism and Magnetic Materials*; Cambridge University Press: New York, NY, USA, 2009.

27. Ermolenko, A. Magnetocrystalline anisotropy of rare earth intermetallics. *IEEE Trans. Magn.* **1976**, *12*, 992–996. https://doi.org/10.1109/TMAG.1976.1059178.

28. Jiang, J.; Bader, S. Rational design of the exchange-spring permanent magnet. *J. Phys. : Condens. Matter* **2014**, *26*, 064214. https://doi.org/10.1088/0953-8984/26/6/064214.

29. Nutor, R.K.; Fan, X.; Ren, S.; Chen, M.; Fang, Y. Research progress of stress-induced magnetic anisotropy in Fe-based amorphous and nanocrystalline alloys. *J. Electromagn. Anal. Appl.* **2017**, *9*, 53. https://doi.org/10.4236/jemaa.2017.94006.

30. Liang, J.; Yue, M.; Zhang, D.; Li, Y.; Xu, X.; Li, H.; Xi, W. Anisotropic SmCo 5 Nanocrystalline Magnet Prepared by Hot Deformation With Bulk Amorphous Precursors. *IEEE Trans. Magn.* **2018**, *54*, 1–4. https://doi.org/10.1109/TMAG.2018.2848281.

31. Schönhöbel, A.M.; Madugundo, R.; Barandiarán, J.M.; Hadjipanayis, G.C.; Palanisamy, D.; Schwarz, T.; Schrefl, T.; Gault, B.; Raabe, D.; Skokov, K.; et al. Nanocrystalline Sm-based 1: 12 magnets. *Acta Mater.* **2020**, *200*, 652–658. https://doi.org/10.1016/j.actamat.2020.08.075.